\documentclass[a4paper,fleqn,final]{cas-sc}

\usepackage[numbers,sort&compress]{natbib}

\usepackage{graphicx}
\usepackage{float}        
\usepackage{placeins}     
\usepackage{caption}

\usepackage{amsmath}
\usepackage{siunitx}

\usepackage{xspace}
\usepackage{microtype}

\def\tsc#1{\csdef{#1}{\textsc{\lowercase{#1}}\xspace}}
\tsc{WGM}
\tsc{QE}

\begin{document}

\let\WriteBookmarks\relax

\def\floatpagepagefraction{1}
\def\textpagefraction{.001}

\shorttitle{Interplay of Composition, Crystallinity, and Chemical Structure in CoHCF and NiHCF Thin Films Prepared at Different Temperatures}

\shortauthors{Garcia et al.}

\title[mode=title]{Interplay of Composition, Crystallinity, and Chemical Structure in CoHCF and NiHCF Thin Films Prepared at Different Temperatures}



\author[1]{Larissa de O. Garcia}
\cormark[1]
\ead{larissa.de.oliveira.garcia@whz.de}
\ead[url]{ORCID: https://orcid.org/0000-0003-0267-6488}

\author[1]{Michael Pohlitz}

\author[2]{Mohammed F. Kalady}

\author[1]{Christian K. Müller}


\affiliation[1]{
organization={Faculty of Physical Engineering/Computer Sciences, University of Applied Sciences Zwickau},
city={Zwickau},
postcode={08056},
country={Germany}
}

\affiliation[2]{
organization={Leibniz Institute for Solid State and Materials Research Dresden (IFW Dresden)},
city={Dresden},
postcode={01069},
country={Germany}
}

\cortext[1]{Corresponding author. Tel.: +49 375 536 1510}

\begin{abstract}
Understanding how growth conditions govern structural order and ion transport in Prussian blue analogues (PBAs) thin films is essential for optimizing their electrochemical performance. Here, cobalt and nickel hexacyanoferrate (CoHCF and NiHCF) thin films were electrodeposited potentiostatically at temperatures between 20 and 60\,°C. A combination of cyclic voltammetry, scanning electron microscopy, X-ray diffraction, and Raman spectroscopy was employed to elucidate the interplay between composition, crystallinity, and chemical structure.

Under identical conditions, CoHCF exhibits a maximum current density approximately 2.2 times higher than NiHCF, indicating significantly faster electrochemical kinetics. X-ray diffraction reveals temperature-dependent lattice expansion without phase transitions, with a maximum near 40\,°C, associated with structural relaxation and compositional variations. Raman spectroscopy further reveals temperature-dependent local structural evolution, where cyanide band narrowing at intermediate temperatures indicates improved short-range order, while band broadening at higher temperatures reflects increased defect density.

These findings demonstrate that temperature-controlled defect redistribution governs both short- and long-range structural order in PBA thin films, directly influencing ion transport and electrochemical response. This work provides new insights into structure–property relationships and establishes deposition temperature as a key parameter for tuning electrochemical functionality in hexacyanoferrate-based electrodes.
\end{abstract}

\maketitle

\section{Introduction}

Prussian blue analogues (PBAs) constitute a versatile class of coordination polymers formed by cyanide-bridged transition-metal ions, exhibiting open framework structures, rich redox chemistry, and remarkable compositional flexibility \cite{Wu2025}. Owing to these characteristics, PBAs have attracted considerable attention for applications in electrochemical energy storage \cite{Javadi2026}, electrocatalysis \cite{Singh2026}, and sensing technologies \cite{Duan2026}. In particular, cobalt- and nickel-based hexacyanoferrates (CoHCF and NiHCF) have emerged as promising candidates due to their chemical stability, tunable electronic properties, strong sensitivity to chemical composition and defect chemistry, making them ideal platforms for investigating structure–property relationships in cyanide-bridged frameworks \cite{Shen2020,Kang2024,Wang2025}.

The functional properties of PBAs are governed by both local and long-range structural order. Parameters such as chemical composition, alkali-metal content, and the concentration of [Fe(CN)$_6$] vacancies directly influence lattice parameters, electronic structure, ionic transport, and vibrational properties \cite{Wu2024,Yin2025}. Additionally, coordinated and interstitial water species affect local bonding environments and charge-compensation mechanisms, thereby influencing redox activity and lattice dynamics. Consequently, even subtle variations in composition and defect chemistry can induce pronounced changes in the functional response of PBAs \cite{Ojwang2021,Sada2024}.

When processed as thin films, these structure–property relationships become more complex. Reduced dimensionality, substrate–film interactions, and kinetically controlled growth conditions can stabilize non-equilibrium structures that differ significantly from their bulk counterparts \cite{Yun2021,Steeger2025}. Electrodeposited Prussian blue and Prussian white thin films have also been investigated for electrochemical switching and device applications, highlighting the sensitivity of their functional response to film growth and electrical transport \cite{Avila2020,Faita2022,Avila2022}. More broadly, electrodeposited Prussian blue architectures have been explored for electrochemical energy-storage applications \cite{Avila2024PB}. Related resistive-switching behavior has also been demonstrated in other functional molecular thin films, providing broader context for the sensitivity of thin-film electrical properties to local structure and morphology \cite{Avila2024Perylene}. As a result, thin-film PBAs often exhibit enhanced defect densities, compositional heterogeneity, and anisotropic structural order, all of which are critical for applications requiring controlled ion transport, stable electrochemical behavior, and well-defined interfacial properties \cite{Nordstrand2021,Avila2025}. Recent variability studies on electrodeposited Prussian blue further emphasize the importance of controlling film formation when interpreting device-level responses \cite{Avila2025}. Despite their technological relevance, systematic studies addressing how synthesis parameters govern defect formation and compositional variation in PBA thin films remain limited.

Among the synthesis parameters, temperature plays a central role in determining crystallinity, morphology, and defect landscapes in PBAs \cite{Li2019}. Electrochemical processing routes have likewise been used to obtain compositionally and structurally controlled functional thin films, illustrating the broader importance of deposition conditions in determining film properties \cite{Quispe2021}. At low temperatures, limited atomic mobility typically results in reduced crystallinity, small grain sizes, and increased structural disorder \cite{Jiang2021}. Increasing temperature enhances atomic diffusion and long-range ordering but may simultaneously promote alkali-ion incorporation, vacancy formation, and the retention of framework water \cite{Hu2009}. This introduces a trade-off between improved crystallinity and defect generation, which remains insufficiently understood in electrodeposited PBA thin films \cite{Camacho2021,Nair2025}.

Defects in PBAs are increasingly recognized not merely as imperfections but as key parameters governing ionic transport and electrochemical response \cite{Iyyappan2026}. Vacancies and local structural distortions can facilitate ion diffusion and modulate redox processes, whereas excessive disorder may compromise structural stability \cite{Janani2025,Sterzinger2025}. Establishing direct correlations between structural evolution and electrochemical behavior is therefore essential for rational materials design. In this context, cyclic voltammetry provides insight into redox activity, charge-transfer kinetics, and ion accessibility in PBA frameworks \cite{Abbaspour2005,Tharuman2025}, while Raman spectroscopy serves as a sensitive probe of local structural order. In particular, the C$\equiv$N stretching modes respond strongly to metal–ligand coordination, oxidation state, and symmetry, enabling the detection of local disorder and compositional variations that may not be captured by diffraction techniques \cite{Weidinger2010,Moretti2018}.

Despite previous studies on the temperature-dependent growth of PBAs in bulk systems, the systematic correlation between deposition temperature, local structural disorder, and electrochemical behavior in electrodeposited thin films remains largely unexplored. The interplay between composition, defect distribution, and multiscale structural ordering, in particular, has not been comprehensively addressed. In this study, we systematically investigate CoHCF and NiHCF thin films electrodeposited at temperatures between 20 and 60\,°C. Using a combination of cyclic voltammetry (CV), X-ray diffraction (XRD), scanning electron microscopy (SEM), and Raman spectroscopy, we establish a direct correlation between deposition temperature, structural evolution, and electrochemical performance. Our results reveal a non-monotonic relationship between crystallinity and electrochemical activity. This demonstrates that defect-mediated structural dynamics play a dominant role in governing ion transport. Furthermore, a comparative analysis of Co- and Ni-based hexacyanoferrates highlights the role of the transition-metal center in modulating structural flexibility and defect sensitivity. These findings provide new insights into temperature-controlled structure–property relationships in PBA thin films and offer practical guidelines for the rational design of cyanide-bridged materials with tailored electrochemical functionality.
\section{Materials and Methods}

\subsection{Sample preparation}
Prussian blue analogues (PBAs) thin films were prepared by potentiostatic electrodeposition using a conventional three-electrode electrochemical cell. A platinum plate served as the counter electrode, a saturated calomel electrode (SCE) as the reference electrode, and Au (50 nm)/Cr (5 nm) layers deposited on Si (100) substrates as the working electrode. Electrochemical deposition was carried out using an Ivium CompactStat potentiostat (Ivium Technologies, Eindhoven, The Netherlands). All potentials reported in this work are referenced to the SCE.

Film growth was performed at a constant applied potential of 0.30 V until a total charge of 50 mC was reached. The effective deposition area (approximately 0.5 cm$^2$) was defined using an adhesive mask applied to the working electrode surface. The electrolyte consisted of an aqueous solution containing 1.0 M KCl, 0.25 mM K$_3$Fe(CN)$_6$, and 0.25 mM metal chloride precursor. The pH was adjusted to approximately 2 by the addition of hydrochloric acid. All chemicals were of analytical grade and used as received. The electrodeposition procedure followed protocols previously established by our group \cite{Pohlitz2022,Garcia2025}.

During deposition, the electrolyte temperature was controlled using a thermostated water bath and continuously monitored with a calibrated thermometer. The deposition temperature was varied between 20 and 60\,°C. This configuration ensured thermal homogeneity of the electrolyte and minimized temperature gradients at the electrode surface during film growth.

Following the nomenclature adopted in our previous work \cite{Garcia2025}, the deposited films were labeled according to the transition metal precursor: CoHCF for cobalt hexacyanoferrate films and NiHCF for nickel hexacyanoferrate films.

\subsection{Material characterization}
Structural characterization was performed by X-ray diffraction (XRD) using an X’Pert MRD diffractometer (PANalytical, Almelo, The Netherlands) with Co K$\alpha$ radiation ($\lambda = 1.78896$~\AA). Measurements were carried out in Bragg--Brentano geometry over a $2\theta$ range of 10--40° with a step size of 0.005°.

Morphological and compositional analyses were conducted using a field-emission scanning electron microscope (FEG-SEM, TESCAN CLARA, Brno, Czech Republic) equipped with an energy-dispersive X-ray spectroscopy (EDS) detector (Ultim Max 65 SDD, Oxford Instruments, Wiesbaden, Germany). SEM imaging and EDS measurements were performed at an accelerating voltage of 20 kV.

Raman spectroscopy was carried out using a confocal Raman system (WITec RISE, Ulm, Germany) with a 532 nm excitation laser. To avoid local heating and sample degradation, the laser power was limited to 0.4 mW for all measurements.

Electrochemical characterization was performed by cyclic voltammetry (CV) using the same three-electrode configuration described above. Measurements were conducted in a 1.0 M KCl aqueous electrolyte within a potential window of 0.0–0.6 V at a scan rate of 100\,mV\,s$^{-1}$. All potentials are reported with respect to the SCE.

\section{Results and Discussion}
\subsection{Electrochemical characterization}
\begin{figure}[t]
\centering
\includegraphics[width=\linewidth]{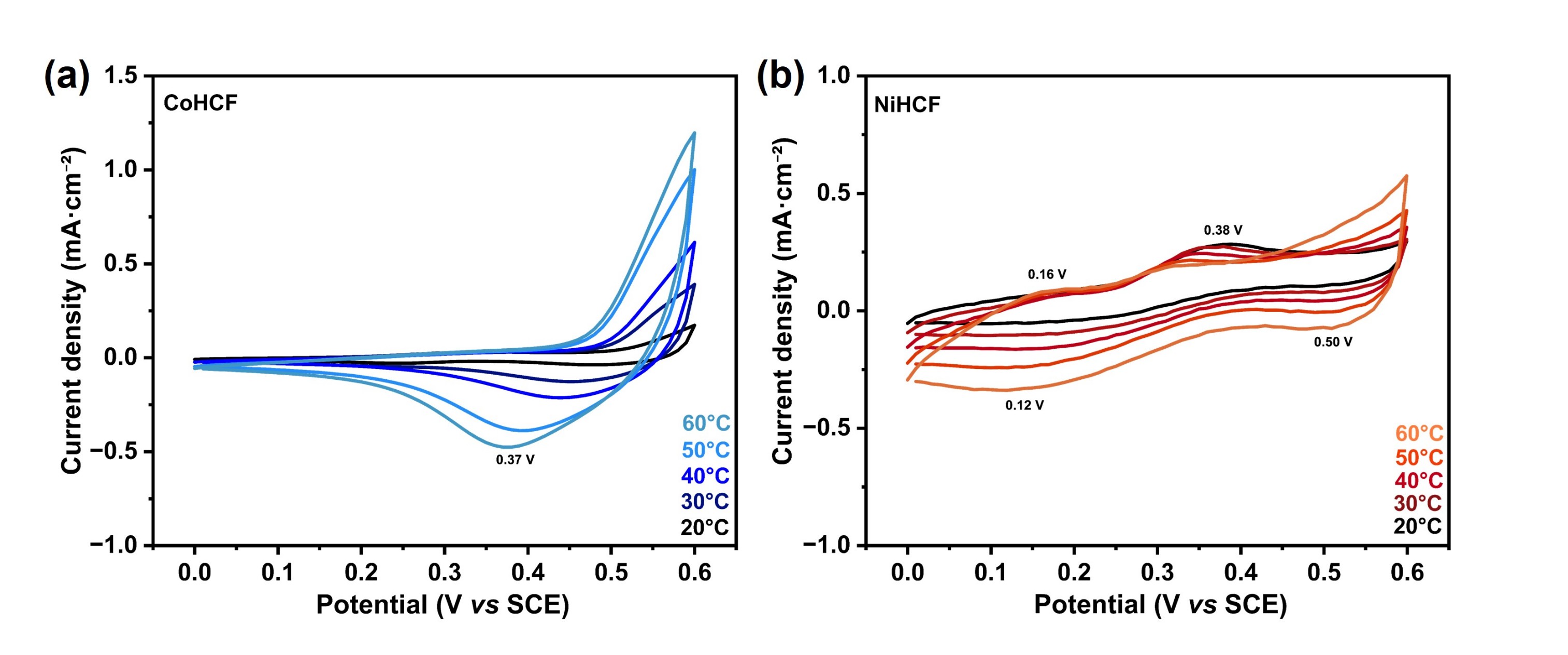}
\caption{Cyclic voltammograms of (a) CoHCF and (b) NiHCF recorded at different temperatures (20–60\,°C) in 1.0\,M KCl aqueous electrolyte at a scan rate of 100\,mV\,s$^{-1}$.}
\label{fig:cv}
\end{figure}

The electrochemical behavior of CoHCF and NiHCF thin films was investigated by cyclic voltammetry (CV) in a 1.0\,M KCl aqueous electrolyte within a potential window of 0.0–0.6\,V vs SCE at temperatures ranging from 20 to 60\,°C (Fig.~\ref{fig:cv}). Electrochemical processing has also been demonstrated as an effective route for preparing functional transition-metal oxide thin films, underscoring the role of processing conditions in controlling electrochemically active film properties \cite{Quispe2021}. This potential range was selected to avoid irreversible oxidation processes and to ensure structural stability during cycling. The upper potential limit was further restricted to isolate the dominant Fe$^{2+}$/Fe$^{3+}$ redox process associated with K$^+$ insertion/extraction, while minimizing contributions from secondary reactions \cite{Garcia2025}.

Both materials exhibit reversible redox features characteristic of Prussian blue analogues (PBAs), in which Fe-centered redox reactions are coupled to alkali-ion transport within an open cyanide-bridged framework \cite{Neff1978,Itaya1982,Kumar2014}. The electrochemical response was highly reproducible across multiple samples.

For CoHCF (Fig.~\ref{fig:cv}a), a well-defined cathodic peak centered at approximately 0.37\,V vs SCE is observed, accompanied by a corresponding anodic peak at higher potentials. Increasing temperature results in a systematic increase in current density and progressive sharpening of the redox peaks. At 60\,°C, CoHCF reaches a maximum cathodic current density of approximately 1.2~mA\,cm$^{-2}$, indicating an enhanced electrochemical response under elevated temperature conditions.

In contrast, NiHCF (Fig.~\ref{fig:cv}b) exhibits broader redox features distributed over a wider potential window (approximately 0.12--0.50\,V vs SCE), indicating the presence of multiple local redox environments rather than a single well-defined transition \cite{Steen2002}. At 60\,°C, the maximum cathodic current density reaches approximately 0.55~mA\,cm$^{-2}$. A quantitative comparison based on the maximum cathodic peak current at 60\,°C shows that CoHCF exhibits a current density approximately 2.2 times higher than NiHCF under identical conditions. This difference provides experimental evidence of enhanced electrochemical kinetics in CoHCF.

The temperature-dependent increase in current density and peak sharpening observed for both materials indicates thermally activated electrochemical behavior, consistent with enhanced ionic mobility and reduced kinetic limitations at elevated temperatures. In CoHCF, the sharper and more intense peaks suggest a more uniform electrochemical response and improved ion accessibility within the framework. In contrast, the broader response observed in NiHCF reflects a distribution of redox sites, typically associated with structural disorder, vacancy distribution, and compositional heterogeneity \cite{Jeerage2002,Xu2018}.

Although electrochemical impedance spectroscopy would be required for a quantitative separation of charge-transfer resistance and diffusion contributions, the consistent trends observed in cyclic voltammetry (CV), together with structural and spectroscopic analyses (scanning electron microscopy (SEM), X-ray diffraction (XRD), and Raman spectroscopy), indicate that the electrochemical response is governed by coupled ionic transport and interfacial processes.

In both systems, the electrochemical behavior is dominated by the reversible Fe$^{2+}$/Fe$^{3+}$ redox couple associated with K$^+$ insertion/extraction, while Co$^{2+}$ and Ni$^{2+}$ primarily act as structural centers without distinct redox activity within the investigated potential window \cite{Neff1978,Itaya1982,Wessells2011}.These electrochemical reactions can be described by the following equilibria for CoHCF \cite{Joseph1991,Kumar2014}:

\begin{equation}
\mathrm{KCo^{II}_{1.5}[Fe^{II}(CN)_6]
\rightleftharpoons
Co^{II}_{1.5}[Fe^{III}(CN)_6] + K^+ + e^-}
\end{equation}

\begin{equation}
\mathrm{K_2Co^{II}[Fe^{II}(CN)_6]
\rightleftharpoons
KCo^{II}[Fe^{III}(CN)_6] + K^+ + e^-}
\end{equation}

An analogous mechanism can be applied to NiHCF, where the Fe$^{2+}$/Fe$^{3+}$ couple similarly dominates the electrochemical response. The coexistence of these configurations is closely linked to the presence of [Fe(CN)$_6$] vacancies and coordinated water molecules, which introduce local structural distortions and a distribution of redox potentials \cite{Buser1977,Jeerage2002,Xu2018}. These effects are consistent with the broader electrochemical features observed, particularly for NiHCF.

A comparison with literature data (Table~\ref{tab:comparison}) indicates that the current densities obtained here fall within the typical range reported for hexacyanoferrate-based thin films. The differences observed between CoHCF and NiHCF are primarily attributed to variations in structural ordering and defect distribution induced by deposition temperature, rather than to significant differences in metal-centered redox activity within the investigated potential window.

Importantly, these results indicate that the distinct electrochemical behavior of CoHCF and NiHCF arises primarily from differences in framework rigidity, defect chemistry, and ion-transport pathways, rather than intrinsic differences in metal-centered redox activity, as discussed in the following sections.

\begin{table}[htbp]
\centering
\small
\caption{Comparison of electrochemical performance of hexacyanoferrate-based materials reported in literature and in this work.}
\label{tab:comparison}

\begin{tabular}{p{2.5cm} p{2.5cm} p{2cm} p{2.2cm} p{4.5cm} p{1.5cm}}
\hline
Material & Synthesis Method & Current Density & Electrolyte/Ion & Key Feature & Ref \\
 & & (mA cm$^{-2}$) & & & \\
\hline
CoHCF (this work) & Electrodeposition & $\sim$1.2 & KCl / K$^+$ 
& Temperature-controlled electrochemical response; maximum at 60\,°C & This work \\
NiHCF (this work) & Electrodeposition & $\sim$0.55 & KCl / K$^+$ 
& Broader redox features; disorder-sensitive response & This work \\
CoHCF & Electrodeposition & 0.5--1.0 & -- / K$^+$ 
& Electrocatalytic enhancement with Au co-deposition & \cite{Kumar2014} \\
NiHCF & Cathodic thin film & 0.2--0.6 & -- / K$^+$ 
& Structural disorder governs electrochemical response & \cite{Steen2002} \\
NiHCF & Chemical (aqueous) & $\sim$0.8 & KNO$_3$ / K$^+$ 
& Improved ion transport in nanostructured films & \cite{Pillai2024} \\
Co/Ni PBA & Composite & 1.0--2.0 & -- 
& Synergistic electrochemical enhancement & \cite{Tharuman2025} \\
\hline
\end{tabular}
\end{table}

\subsection{Morphological characterization by SEM and EDS}

\begin{figure}[t]
    \centering
    \includegraphics[width=\linewidth]{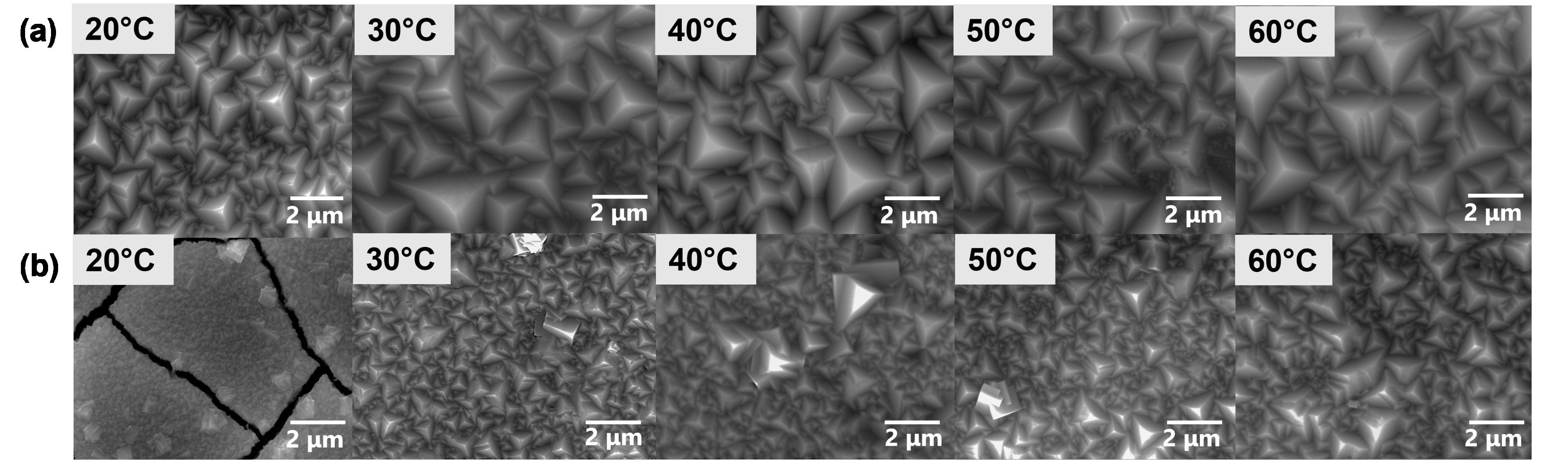}
    \caption{SEM images of (a) CoHCF and (b) NiHCF thin films electrodeposited at different temperatures (20--60\,°C).}
    \label{fig:SEM}
\end{figure}

To elucidate the structural origins of the distinct electrochemical behavior observed for CoHCF and NiHCF, their morphology and microstructure were investigated by scanning electron microscopy (SEM) (Fig.~\ref{fig:SEM}). As electrodeposition was performed within the electrochemically active Fe$^{2+}$/Fe$^{3+}$ redox regime, the resulting morphologies reflect kinetically controlled growth conditions governed by the interplay between nucleation and crystal growth.

For both materials, a clear evolution of surface morphology with deposition temperature is observed, highlighting the role of temperature in governing growth processes. Increasing temperature enhances surface diffusion and reduces kinetic barriers, thereby promoting crystallite growth and coalescence \cite{Yang2016,Liu2024}.

However, the extent of this effect differs significantly between the two systems. CoHCF films (Fig.~\ref{fig:SEM}a) exhibit relatively large and well-defined faceted crystallites across the entire temperature range, indicating a predominantly growth-controlled deposition regime. Only minor morphological variations are observed between 20\,°C and 60\,°C. At 20\,°C, the films already display well-developed faceted grains, while at 60\,°C a slight increase in crystallite size and improved inter-grain connectivity are observed.

This comparatively weak temperature dependence suggests a kinetically favorable growth regime that is less sensitive to thermal variations. Such behavior is consistent with stronger metal--ligand interactions and higher framework rigidity in Co-based PBAs, which favor structural stability and long-range ordering \cite{Jeerage2002,Xu2018}.

Although no statistical grain size analysis was performed, the observed morphological features were consistent across multiple regions and independently prepared samples, supporting the reproducibility of the trends discussed.

In contrast with CoHCF, NiHCF films (Fig.~\ref{fig:SEM}b) exhibit a pronounced temperature dependence. At 20\,°C, the films display an inhomogeneous morphology with visible cracks and poorly connected grains, indicative of limited surface diffusion and nucleation-dominated growth. At intermediate temperatures (30--40\,°C), the morphology evolves into dense and more homogeneous granular structures, reflecting increased nucleation density and enhanced surface mobility. At higher temperatures (50--60\,°C), significant grain growth and coalescence occur, leading to larger aggregates and improved film continuity. This transition from nucleation-limited to growth-dominated regimes highlights the stronger sensitivity of NiHCF to deposition temperature. The improved homogeneity observed at intermediate temperatures is consistent with the reduced electrochemical polarization and enhanced current response observed in CV measurements.

\begin{figure}[t]
    \centering
    \includegraphics[width=\linewidth]{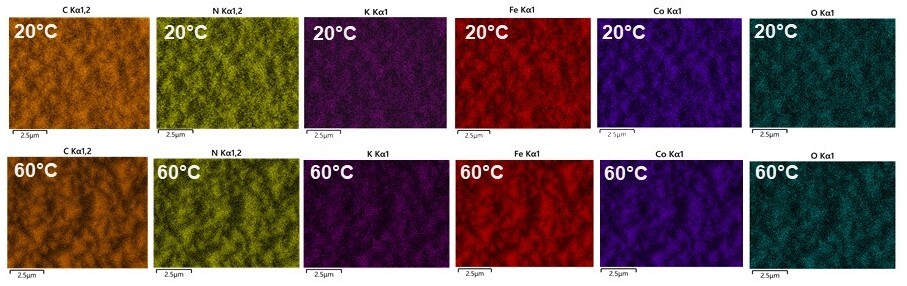}
    \caption{EDX elemental mapping of CoHCF thin films electrodeposited at 20 and 60\,°C.}
    \label{fig:EDX-Co}
\end{figure}

Elemental composition and spatial distribution were further examined by energy-dispersive X-ray spectroscopy (EDS). Representative elemental maps for films deposited at 20 and 60\,°C are shown in Figs.~\ref{fig:EDX-Co} and \ref{fig:EDX-Ni}, while additional maps for intermediate temperatures are provided in the Supplementary Information. For both materials and all investigated temperatures, the EDS maps confirm a homogeneous spatial distribution of the constituent elements, with no evidence of segregation or secondary phase formation.

\begin{figure}[t]
    \centering
    \includegraphics[width=\linewidth]{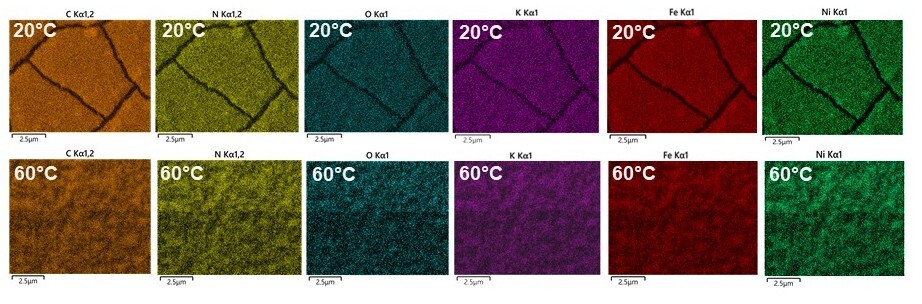}
    \caption{EDX elemental mapping of NiHCF thin films electrodeposited at 20 and 60\,°C.}
    \label{fig:EDX-Ni}
\end{figure}

Semi-quantitative EDS analysis of CoHCF and NiHCF films (Tables~\ref{tab:eds_cohcf} and \ref{tab:eds_nihcf}) confirms the presence of potassium and oxygen, reflecting the incorporation of $K^+$ ions for charge compensation and the presence of coordinated and/or interstitial water molecules. 

It should be noted that EDS measurements probe near-surface composition and may be influenced by interaction volume effects as well as possible substrate contributions. In addition, absolute atomic percentages should be interpreted with caution due to the limitations of EDS in light-element quantification; however, the observed trends provide reliable comparative information across the temperature series.

A systematic dependence of the O/K ratio on deposition temperature is observed for both materials. In general, increasing temperature promotes higher potassium incorporation and a relative decrease in oxygen content. For CoHCF, the K content increases from 7.36 at.\% at 20\,°C to 9.82 at.\% at 60\,°C, while the O content decreases from 6.82 to 4.82 at.\%. A similar trend is observed for NiHCF, where the K content increases from 5.30 to 8.22 at.\%, whereas the O content varies within a narrower range between 4.27 and 5.05 at.\%.

These trends suggest temperature-dependent changes in the balance between alkali-ion incorporation and hydration within the framework. Although EDS does not directly quantify vacancy concentration, the observed variations are consistent with modifications in defect distribution and coordinated water content.

Independent evidence of the chemical states and composition of CoHCF and NiHCF thin films has been reported in our previous work \cite{Garcia2025}, where X-ray photoelectron spectroscopy (XPS) confirmed the presence of Co$^{2+}$, Ni$^{2+}$, and mixed Fe$^{2+}$/Fe$^{3+}$ oxidation states. These results support the compositional interpretation presented here and further validate the use of EDS for comparative analysis across the temperature series.

Such compositional changes are expected to influence both lattice parameters and local structural environments, in agreement with the XRD and Raman results. Increased potassium incorporation may partially compensate framework vacancies, while variations in oxygen content are associated with changes in coordinated and/or interstitial water. Together, these factors modify the local bonding environment and contribute to the structural stability of the PBA framework.

Taken together, the combined morphological and compositional analyses reveal distinct growth mechanisms for CoHCF and NiHCF under identical deposition conditions. CoHCF exhibits a relatively stable, growth-dominated morphology with weak temperature dependence, whereas NiHCF undergoes a clear transition from nucleation-controlled to growth-controlled regimes as temperature increases. These differences influence structural ordering and defect distribution, providing a consistent and unified interpretation of the temperature-dependent trends observed across XRD, Raman, and electrochemical measurements.

\begin{table}
\centering
\caption{Elemental composition (at.\%) of CoHCF thin films obtained by EDS mapping at different deposition temperatures.}
\label{tab:eds_cohcf}
\begin{tabular}{c c c c c c c}
\hline
Temperature (°C) & C & N & K & O & Fe & Co \\
\hline
20 & 41.12 & 36.01 & 7.36 & 6.82 & 4.22 & 4.46 \\
30 & 37.84 & 35.14 & 8.69 & 8.04 & 5.00 & 5.28 \\
40 & 39.38 & 35.25 & 9.36 & 5.78 & 5.08 & 5.15 \\
50 & 39.56 & 35.88 & 9.32 & 5.07 & 5.13 & 5.05 \\
60 & 39.45 & 35.36 & 9.82 & 4.82 & 5.42 & 5.13 \\
\hline
\end{tabular}
\end{table}

\begin{table}
\centering
\caption{Elemental composition (at.\%) of NiHCF thin films obtained by EDS mapping at different deposition temperatures.}
\label{tab:eds_nihcf}
\begin{tabular}{c c c c c c c}
\hline
Temperature (°C) & C & N & K & O & Fe & Ni \\
\hline
20 & 43.77 & 39.60 & 5.30 & 4.27 & 3.31 & 3.75 \\
30 & 41.87 & 37.57 & 7.23 & 4.90 & 4.32 & 4.11\\
40 & 41.73 & 37.44 & 7.59 & 4.51 & 4.34 & 4.39 \\
50 & 40.41 & 37.94 & 7.20 & 5.72 & 4.98 & 3.74\\
60 & 40.20 & 37.47 & 8.22 & 5.05 & 4.80 & 4.27 \\
\hline
\end{tabular}
\end{table}

\subsection{Structural characterization by X-ray diffraction}

X-ray diffraction (XRD) was employed to investigate the crystallographic structure and lattice evolution of CoHCF and NiHCF thin films as a function of deposition temperature. In Prussian blue analogues (PBAs), the lattice structure is highly sensitive to alkali-ion occupancy, [Fe(CN)$_6$] vacancies, and local structural distortions within the metal–cyanide framework. These factors can induce measurable deviations from ideal cubic symmetry and significantly influence long-range structural order \cite{Buser1977,Jeerage2002,Xu2018}. Consequently, the temperature-dependent compositional variations identified by EDS are expected to impact both lattice parameters and structural ordering.

\begin{figure*}[t]
\centering
\includegraphics[width=1\textwidth]{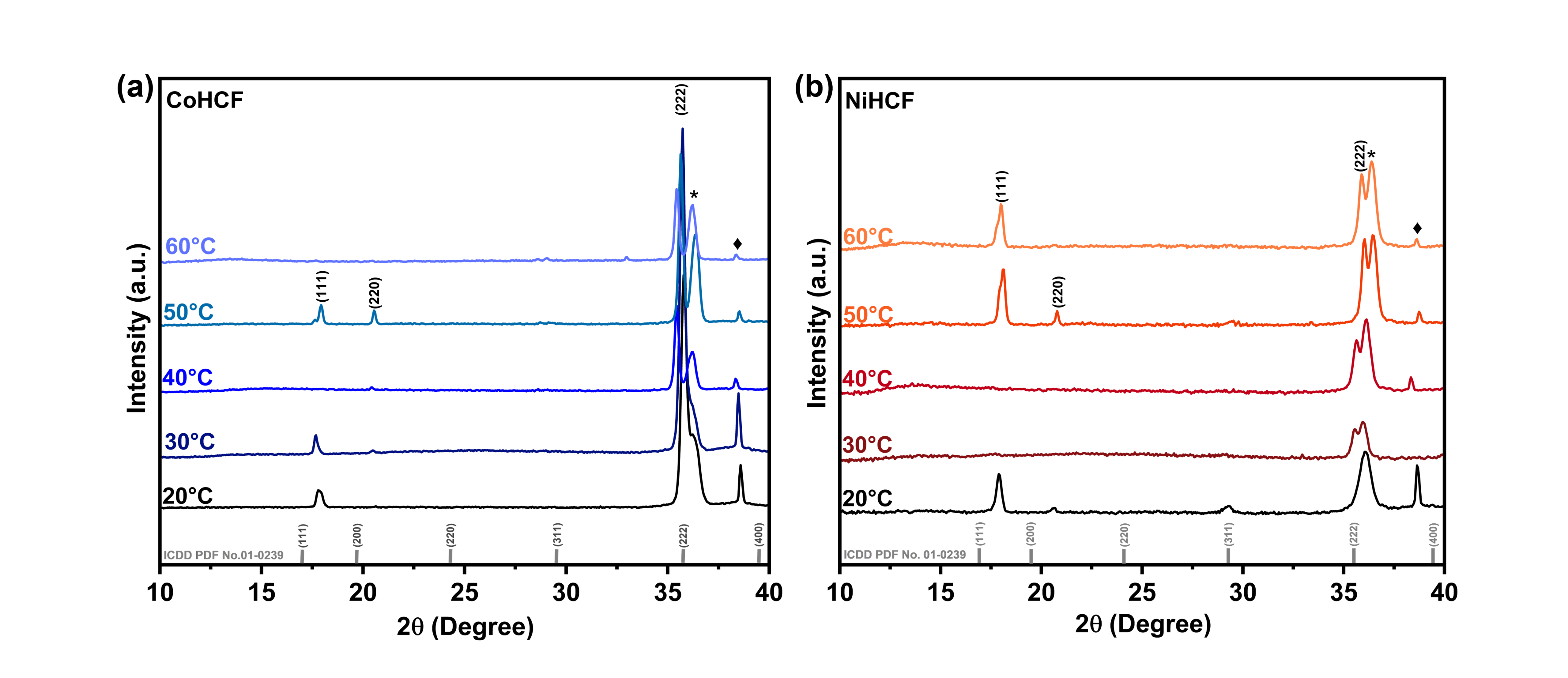}
\caption{
X-ray diffraction (XRD) patterns of (a) CoHCF and (b) NiHCF thin films electrodeposited at different temperatures (20--60\,°C). Reflections marked with asterisks (*) are associated with structural disorder, while diamond symbols ($\blacklozenge$) indicate the presence of a surface oxide layer. Reference stick patterns corresponding to the cubic Prussian blue analogue structure (space group $Fm\bar{3}m$, ICDD PDF No.~01-0239) were adapted to Co K$\alpha$ radiation and are shown at the bottom of each panel for comparison.
}
\label{fig:XRD}
\end{figure*}

The XRD patterns collected in the 2$\theta$ range of 10–40° are shown in Fig.~\ref{fig:XRD}. This angular range emphasizes the characteristic reflections of the PBA framework while minimizing contributions from the Au/Cr substrate, whose most intense peaks occur at higher diffraction angles \cite{Waseda1980,Cullity2001}. The dominant reflections observed at approximately 17°, 24°, 35°, and 39° correspond to the (111), (220), (222), and (400) planes, respectively, consistent with a face-centered cubic structure (space group $Fm\bar{3}m$).

To facilitate phase identification, reference stick patterns corresponding to the cubic PBA structure were included in Fig.~\ref{fig:XRD}, based on standard diffraction data from the ICDD database (PDF No.~01-0239). The reference data were adapted to Co K$\alpha$ radiation to match the experimental conditions. Although the original ICDD data are typically reported for Cu K$\alpha$ radiation, this conversion preserves the relative peak positions and crystallographic indexing, enabling direct comparison with the experimental patterns.

A clear evolution of peak intensity and sharpness with deposition temperature is observed for both materials. For CoHCF (Fig.~\ref{fig:XRD}a), relatively well-defined reflections are obtained at intermediate temperatures, while partial peak broadening and intensity reduction occur at 40 and 60\,°C. Similarly, NiHCF (Fig.~\ref{fig:XRD}b) exhibits broader and less intense reflections, particularly at 30 and 40\,°C. These features indicate a reduction in long-range structural coherence and an increase in structural disorder.

The observed peak broadening and intensity variations are attributed to temperature-dependent structural disorder, including defect formation, lattice strain, and finite crystallite size effects, rather than to the formation of secondary crystalline phases. This interpretation is supported by the absence of additional diffraction peaks and the good agreement between experimental reflections and the reference stick pattern. Such behavior is commonly reported for electrodeposited PBA thin films, where kinetic growth conditions promote non-equilibrium structures with varying vacancy concentrations and compositional heterogeneity \cite{Widmann2002,Warren1969}.

Due to the relatively low peak-to-background ratio and the intrinsic limitations of thin-film diffraction data, full Rietveld refinement was not performed \cite{Young1993}. Instead, lattice parameters were estimated from the position of the (222) reflection assuming cubic symmetry ($Fm\bar{3}m$), which provides the highest intensity and most reliable peak position across all samples. Although preferred orientation effects cannot be completely excluded, the consistent evolution of the (222) peak across the temperature series enables a reliable comparative analysis of lattice trends. This approach is commonly adopted for thin-film PBAs when peak broadening and substrate contributions limit full-pattern refinement.

The calculated lattice parameters are summarized in Table~\ref{tab:lattice_parameters}. For CoHCF, the lattice parameter increases from approximately 10.08\,Å at 20\,°C to a maximum of 10.18\,Å at 40\,°C, followed by a slight decrease at higher temperatures. This non-monotonic behavior indicates temperature-induced structural relaxation and framework rearrangement.

In contrast, NiHCF exhibits slightly smaller lattice parameters (10.01–10.14\,Å), consistent with the smaller ionic radius of Ni compared to Co \cite{Lee2024}. A similar non-monotonic trend is observed, with a maximum at 30\,°C, suggesting that deposition temperature primarily influences defect distribution and local structural ordering rather than simple thermal expansion.

These results demonstrate that deposition temperature governs the balance between structural ordering and defect formation in PBA thin films. At intermediate temperatures, enhanced atomic mobility promotes structural rearrangement and lattice expansion, whereas higher temperatures favor defect incorporation and partial structural relaxation. The absence of peak splitting or additional reflections confirms that the cubic $Fm\bar{3}m$ structure is preserved across the entire temperature range.

When considered together with SEM and Raman analyses, the XRD results indicate that temperature-dependent compositional variations and defect redistribution control both long-range structural coherence and local structural disorder, which ultimately govern the electrochemical response of CoHCF and NiHCF thin films.

\begin{table}[h]
\centering
\caption{Lattice parameters of CoHCF and NiHCF thin films as a function of deposition temperature, estimated from the (222) reflection.}
\label{tab:lattice_parameters}
\begin{tabular}{c cc cc}
\hline
\multirow{2}{*}{Temperature (°C)} 
& \multicolumn{2}{c}{CoHCF} 
& \multicolumn{2}{c}{NiHCF} \\
\cline{2-5}
 & $2\theta_{(222)}$ (°) & $a$ (Å) 
 & $2\theta_{(222)}$ (°) & $a$ (Å) \\
\hline
20 & 35.81 & 10.08 & 36.04 & 10.01 \\
30 & 35.73 & 10.10 & 35.57 & 10.14 \\
40 & 35.45 & 10.18 & 35.62 & 10.13 \\
50 & 35.62 & 10.13 & 36.04 & 10.01 \\
60 & 35.46 & 10.17 & 35.89 & 10.05 \\
\hline
\end{tabular}

\end{table}

\subsection{Raman Spectroscopy}

Raman spectroscopy was employed to investigate the temperature-dependent structural evolution of CoHCF and NiHCF thin films, with particular emphasis on the cyanide stretching region (2000--2250\,cm$^{-1}$). This spectral range is highly sensitive to local bonding environments, oxidation states, and defect chemistry in Prussian blue analogues (PBAs) \cite{Kettle2011}. The $C\equiv N$ stretching frequency is governed by the $\sigma$-donor and $\pi$-acceptor character of the metal ions coordinated to the carbon and nitrogen termini; consequently, variations in local symmetry and compositional environment induce measurable frequency shifts \cite{Nakamoto2009}. Owing to this strong coupling, Raman spectroscopy provides a sensitive probe of short-range order and local structural heterogeneities—such as $[Fe(CN)_6]$ vacancies and interactions with coordinated water—which are often not detectable by diffraction techniques probing long-range periodicity. In this work, the analysis focuses on qualitative trends in band position and linewidth, which serve as indicators of temperature-dependent local structural disorder.

\begin{figure}[t] 
\centering \includegraphics[width=\linewidth]{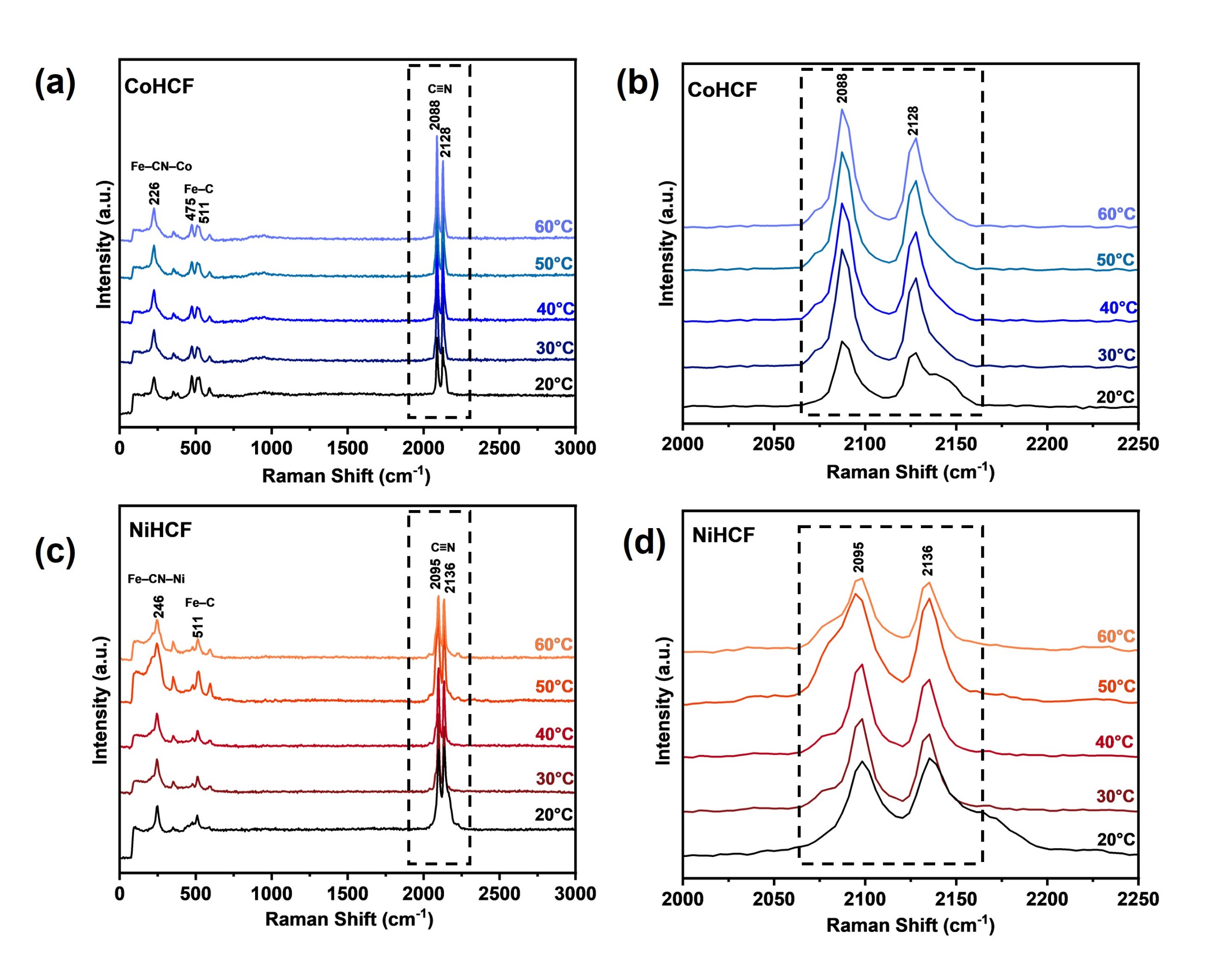}
\caption{Raman spectra of (a) CoHCF and (b) NiHCF thin films deposited at different temperatures. Panels (a,c) show the full spectral range, and panels (b,d) present the enlarged CN stretching region (2000--2250\,cm$^{-1}$).}
\label{fig:raman} 
\end{figure}

The most intense features appear between approximately 2000 and 2150\,cm$^{-1}$, corresponding to the $\nu(C\equiv N)$ stretching modes of $Fe-CN-M$ linkages. The presence of a well-defined doublet (e.g., at 2088 and 2128\,cm$^{-1}$ for CoHCF, Fig.~\ref{fig:raman}a, and 2095 and 2136\,cm$^{-1}$ for NiHCF, Fig.~\ref{fig:raman}c) reflects the coexistence of distinct local coordination environments. These arise from variations in metal–cyanide bonding, local compositional heterogeneity, and electrostatic interactions within the PBA lattice \cite{Kettle2011}. In both systems, the low-frequency region ($<$600\,cm$^{-1}$) is dominated by vibrational modes associated with $Fe-C$ stretching and $Fe-CN-M$ ($M = Co, Ni$) bending, confirming the preservation of the cyanide-bridged framework.

At low deposition temperatures (20-- 30\,°C), both CoHCF (Fig.~\ref{fig:raman}b) and NiHCF (Fig.~\ref{fig:raman}d) films exhibit pronounced band broadening across the spectral range. This behavior indicates increased local structural disorder and reduced short-range ordering, characteristic of films formed under kinetically limited conditions. Restricted atomic mobility at low temperatures promotes rapid nucleation and the formation of small coherent domains, resulting in a distribution of $Fe-CN-M$ bond lengths and angles. This interpretation is consistent with SEM observations of heterogeneous morphology and XRD results indicating reduced structural coherence.

As the deposition temperature increases to 40\,°C, a noticeable narrowing of the Raman bands is observed, particularly in the $C\equiv N$ stretching region. This trend is consistent with the formation of a more homogeneous cyanide framework with improved short-range ordering, driven by enhanced atomic mobility and more efficient structural rearrangement during film growth. Notably, this temperature coincides with the maximum lattice parameter obtained from XRD and the highest electrochemical response observed in CV measurements. These combined observations suggest that 40\,°C represents a structural–kinetic optimum, where the expanded lattice framework is associated with improved local ordering and enhanced ion-transport pathways.

At higher deposition temperatures (50-- 60\,°C), the Raman spectra show a subtle but systematic re-broadening of the $C\equiv N$ stretching bands, particularly at the band edges, despite the absence of any crystallographic phase transition. This behavior indicates the emergence of additional local structural heterogeneities rather than a loss of long-range order. Elevated temperatures promote compositional fluctuations, alkali-ion incorporation, and defect redistribution, which can modify the local electrostatic environment of the cyanide ligands through mechanisms such as changes in hydrogen-bonding interactions, local symmetry distortions near defect sites, and strain fields associated with lattice rearrangements \cite{Nielsen2026}. This trend is consistent with the temperature-dependent variations in the O/K ratio observed by EDS, supporting the role of compositional changes in governing local structure.

A comparison between CoHCF and NiHCF reveals that NiHCF exhibits more pronounced band broadening at elevated temperatures, indicating greater sensitivity to local structural perturbations. This behavior is consistent with differences in metal–nitrogen bonding strength and electronic configuration between $Co^{2+}$ and $Ni^{2+}$, which influence the framework's tolerance to defects and compositional variations \cite{Xu2018,Garcia2025,Jain2024}.

Although detailed peak deconvolution is beyond the scope of the present work, the systematic evolution of band linewidths provides robust qualitative evidence of temperature-dependent structural disorder. Overall, these results indicate that the structural evolution of CoHCF and NiHCF thin films is governed by a balance between long-range structural ordering and defect-induced local disorder. Specifically, low deposition temperatures favor heterogeneous frameworks with limited short-range order, whereas intermediate temperatures (40\,°C) promote a more homogeneous cyanide network. In contrast, higher temperatures introduce composition- and defect-related distortions that broaden Raman features, despite the improved average structural ordering observed by XRD.

These local structural variations correlate with the temperature-dependent electrochemical behavior observed in CV measurements, indicating that composition- and defect-mediated structural dynamics play a key role in governing electrochemical performance in Prussian blue analogue thin films.

\section{Conclusion}

This work demonstrates that deposition temperature plays a critical role in balancing structural ordering and defect formation in Prussian blue analogues (PBAs) thin films. A clear non-monotonic relationship is observed between structural evolution and electrochemical performance, indicating that optimal ion transport is achieved by balancing framework ordering and defect-mediated transport pathways.

Quantitative comparisons based on cyclic voltammetry reveal that, under identical conditions, CoHCF exhibits approximately 2.2 times higher current densities than NiHCF, providing direct evidence of faster electrochemical kinetics. This enhanced performance is attributed to CoHCF's higher framework rigidity and lower sensitivity to defect-induced disorder.

Combined scanning electron microscopy (SEM), energy dispersive spectroscopy (EDS), X-ray diffraction (XRD), and Raman analyses demonstrate that deposition temperature modulates both long-range crystallinity and local structural environments through changes in composition, defect distribution, and hydration state. Specifically, intermediate temperatures promote improved structural homogeneity, whereas higher temperatures favor increased defect incorporation and local disorder.

These results show that the electrochemical performance of PBA thin films is not only governed by crystallinity, but also by the complex interplay of defect chemistry, lattice dynamics, and ion transport pathways. Therefore, deposition temperature emerges as a key and accessible parameter for controlling these properties.

This study provides new insights into the temperature-dependent structure-property relationships of electrodeposited PBAs and establishes a practical framework for the rational design and optimization of hexacyanoferrate-based thin-film electrodes. Notably, the results demonstrate that defect engineering, controlled by deposition temperature, is an effective method for tailoring ion transport and electrochemical functionality in cyanide-bridged materials.




\section*{CRediT authorship contribution statement}
L. O. Garcia: Investigation, Writing – original draft.
M. Pohlitz: Methodology.
M. F. Kalady: Methodology.
C. K. Müller: Conceptualization, Supervision, Funding acquisition.

\section*{Declaration of competing interest}
The authors declare that they have no known competing financial interests.

\section*{Data availability}
The data that support the findings of this study are available from the corresponding author upon reasonable request.

\section*{Acknowledgments}
The authors thank Birgit Opitz from IFW Dresden for support with XRD measurements.

\section*{Funding}
This work was supported by the Deutsche Forschungsgemeinschaft (DFG, Project No. 531524052) and by the European Social Fund (ESF) and the Free State of Saxony within the Landesinnovationsförderprogramm (Application No. 100670500).

\section*{Appendix A. Supplementary data}
Supplementary data associated with this article can be found in the online version.

\bibliographystyle{cas-model2-names}
\bibliography{cas-refs}

@article{Garcia2025,
  title = {Electrodeposited Co and Ni Hexacyanoferrates: Insights into Structure and Morphology},
  volume = {18},
  ISSN = {1996-1944},
  url = {http://dx.doi.org/10.3390/ma18245547},
  DOI = {10.3390/ma18245547},
  number = {24},
  journal = {Materials},
  publisher = {MDPI AG},
  author = {Garcia,  Larissa de O. and Pohlitz,  Michael and Kalady,  Mohammed F. and R\"{o}der,  Falk and Lubk,  Axel and Wolf,  Daniel and M\"{u}ller,  Christian K.},
  year = {2025},
  month = dec,
  pages = {5547}
}

@article{Kumar2014,
  title = {Influence of co-electrodeposited Gold particles on the electrocatalytic properties of CoHCF thin films},
  volume = {139},
  ISSN = {0013-4686},
  url = {http://dx.doi.org/10.1016/j.electacta.2014.06.156},
  DOI = {10.1016/j.electacta.2014.06.156},
  journal = {Electrochimica Acta},
  publisher = {Elsevier BV},
  author = {Kumar,  Alam Venugopal Narendra and Joseph,  James},
  year = {2014},
  month = sep,
  pages = {88--95}
}

@article{Joseph1991,
  title = {Electrodes modified with cobalt hexacyanoferrate},
  volume = {304},
  ISSN = {0022-0728},
  url = {http://dx.doi.org/10.1016/0022-0728(91)85509-N},
  DOI = {10.1016/0022-0728(91)85509-n},
  number = {1--2},
  journal = {Journal of Electroanalytical Chemistry and Interfacial Electrochemistry},
  publisher = {Elsevier BV},
  author = {Joseph,  James and Gomathi,  H. and Rao,  G.Prabhakara},
  year = {1991},
  month = apr,
  pages = {263--269}
}

@article{Wu2025,
  title = {Recent Progress in Prussian Blue Analog Nanomaterials: Structural Engineering and Functional Applications},
  volume = {19},
  ISSN = {1864-564X},
  url = {http://dx.doi.org/10.1002/cssc.202501886},
  DOI = {10.1002/cssc.202501886},
  number = {1},
  journal = {ChemSusChem},
  publisher = {Wiley},
  author = {Wu,  Ke and Zhang,  Guangxun and Yan,  Bingyi and Pang,  Huan},
  year = {2025},
  month = nov 
}

@article{Wang2025,
  title = {Improving Electrochemical Performance of Cobalt Hexacyanoferrate as Magnesium Ion Battery Cathode Material by Nickel Doping},
  volume = {11},
  ISSN = {2313-0105},
  url = {http://dx.doi.org/10.3390/batteries11060213},
  DOI = {10.3390/batteries11060213},
  number = {6},
  journal = {Batteries},
  publisher = {MDPI AG},
  author = {Wang,  Jinxing and Zhang,  Peiyang and Wang,  Jiaxu and Huang,  Guangsheng and Wang,  Jingfeng and Pan,  Fusheng},
  year = {2025},
  month = may,
  pages = {213}
}

@article{Kang2024,
  title = {Enhancing Rate Performance and Stability of Nickel Hexacyanoferrate through K+ Substitution for Na+ Guest Ions},
  volume = {MA2024-02},
  ISSN = {2151-2043},
  url = {http://dx.doi.org/10.1149/MA2024-0291297mtgabs},
  DOI = {10.1149/ma2024-0291297mtgabs},
  number = {9},
  journal = {ECS Meeting Abstracts},
  publisher = {The Electrochemical Society},
  author = {Kang,  San Chun and Song,  Na Young and Chun,  Sang-Eun},
  year = {2024},
  month = nov,
  pages = {1297--1297}
}

@article{Shen2020,
  title = {High-stability monoclinic nickel hexacyanoferrate cathode materials for ultrafast aqueous sodium ion battery},
  volume = {388},
  ISSN = {1385-8947},
  url = {http://dx.doi.org/10.1016/j.cej.2020.124228},
  DOI = {10.1016/j.cej.2020.124228},
  journal = {Chemical Engineering Journal},
  publisher = {Elsevier BV},
  author = {Shen,  Liuxue and Jiang,  Yu and Liu,  Yuefeng and Ma,  Junlin and Sun,  Tongrui and Zhu,  Nan},
  year = {2020},
  month = may,
  pages = {124228}
}

@article{Li2019,
  title = {Chemical Properties,  Structural Properties,  and Energy Storage Applications of Prussian Blue Analogues},
  volume = {15},
  ISSN = {1613-6829},
  url = {http://dx.doi.org/10.1002/smll.201900470},
  DOI = {10.1002/smll.201900470},
  number = {32},
  journal = {Small},
  publisher = {Wiley},
  author = {Li,  Wei-Jie and Han,  Chao and Cheng,  Gang and Chou,  Shu-Lei and Liu,  Hua-Kun and Dou,  Shi-Xue},
  year = {2019},
  month = apr 
}

@article{Jiang2021,
  title = {Room temperature synthesis of high-entropy Prussian blue analogues},
  volume = {79},
  ISSN = {2211-2855},
  url = {http://dx.doi.org/10.1016/j.nanoen.2020.105464},
  DOI = {10.1016/j.nanoen.2020.105464},
  journal = {Nano Energy},
  publisher = {Elsevier BV},
  author = {Jiang,  Wei and Wang,  Tao and Chen,  Hao and Suo,  Xian and Liang,  Jiyuan and Zhu,  Wenshuai and Li,  Huaming and Dai,  Sheng},
  year = {2021},
  month = jan,
  pages = {105464}
}

@article{Hu2009,
  title = {Prussian Blue mesocrystals prepared by a facile hydrothermal method},
  volume = {11},
  ISSN = {1466-8033},
  url = {http://dx.doi.org/10.1039/b911613n},
  DOI = {10.1039/b911613n},
  number = {11},
  journal = {CrystEngComm},
  publisher = {Royal Society of Chemistry (RSC)},
  author = {Hu,  Ming and Jiang,  Ji-Sen and Ji,  Rui-Ping and Zeng,  Yi},
  year = {2009},
  pages = {2257}
}

@article{Camacho2021,
  title = {Impact of Synthesis Conditions in Na-Rich Prussian Blue Analogues},
  volume = {13},
  ISSN = {1944-8252},
  url = {http://dx.doi.org/10.1021/acsami.1c09378},
  DOI = {10.1021/acsami.1c09378},
  number = {36},
  journal = {ACS Applied Materials \& Interfaces},
  publisher = {American Chemical Society (ACS)},
  author = {Camacho,  Paula Sanz and Wernert,  Romain and Duttine,  Mathieu and Wattiaux,  Alain and Rudola,  Ashish and Balaya,  Palani and Fauth,  Fran\c{c}ois and Berthelot,  Romain and Monconduit,  Laure and Carlier,  Dany and Croguennec,  Laurence},
  year = {2021},
  month = sep,
  pages = {42682--42692}
}

@article{Nair2025,
  title = {The impact of synthetic modifications on FeCo Prussian blue analogue and their consequential effects on the Fenton degradation of acetaminophen},
  volume = {13},
  ISSN = {2213-3437},
  url = {http://dx.doi.org/10.1016/j.jece.2025.115690},
  DOI = {10.1016/j.jece.2025.115690},
  number = {2},
  journal = {Journal of Environmental Chemical Engineering},
  publisher = {Elsevier BV},
  author = {Nair,  Keerthi M. and Thomas,  Nishanth and Pallilavalappil,  Sreedhanya and Hinder,  Steven J. and Brennan,  Barry and Pillai,  Suresh C.},
  year = {2025},
  month = apr,
  pages = {115690}
}

@article{Yin2025,
  title = {Unconventional hexagonal open Prussian blue analog structures},
  volume = {16},
  ISSN = {2041-1723},
  url = {http://dx.doi.org/10.1038/s41467-024-55775-w},
  DOI = {10.1038/s41467-024-55775-w},
  number = {1},
  journal = {Nature Communications},
  publisher = {Springer Science and Business Media LLC},
  author = {Yin,  Jinwen and Wang,  Jing and Sun,  Mingzi and Yang,  Yajie and Lyu,  Jia and Wang,  Lei and Dong,  Xinglong and Ye,  Chenliang and Bao,  Haibo and Guo,  Jun and Chen,  Bo and Zhou,  Xichen and Zhai,  Li and Li,  Zijian and He,  Zhen and Luo,  Qinxin and Meng,  Xiang and Ma,  Yangbo and Zhou,  Jingwen and Lu,  Pengyi and Wang,  Yunhao and Niu,  Wenxin and Zheng,  Zijian and Han,  Yu and Zhang,  Daliang and Xi,  Shibo and Yuan,  Ye and Huang,  Bolong and Guo,  Peng and Fan,  Zhanxi},
  year = {2025},
  month = jan 
}

@article{Sterzinger2025,
  title = {Degradation Mechanisms of Prussian Blue Analogues and State-of-the-Art Approaches for Stability Optimization: A Review},
  volume = {129},
  ISSN = {1932-7455},
  url = {http://dx.doi.org/10.1021/acs.jpcc.5c00877},
  DOI = {10.1021/acs.jpcc.5c00877},
  number = {15},
  journal = {The Journal of Physical Chemistry C},
  publisher = {American Chemical Society (ACS)},
  author = {Sterzinger,  Johannes and Streng,  Raphael and Chen,  Shuai and G\"{o}tz,  Rainer and Hu,  Wang and Li,  Jinyang and Bandarenka,  Aliaksandr S.},
  year = {2025},
  month = apr,
  pages = {7135--7153}
}

@article{Janani2025,
  title = {Strategic insights into Prussian Blue Analogues-based catalysts: Design and regulation for enhanced electrochemical energy storage and conversion},
  volume = {9},
  ISSN = {2949-8228},
  url = {http://dx.doi.org/10.1016/j.nxmate.2025.100930},
  DOI = {10.1016/j.nxmate.2025.100930},
  journal = {Next Materials},
  publisher = {Elsevier BV},
  author = {Janani,  Gnanaprakasam and Park,  Soobin and Surendran,  Subramani and Lim,  Yoongu and Moon,  Dae Jun and Jeong,  Gyoung Hwa and Choi,  Heechae and Kwon,  Gibum and Lu,  Xiaoyan and Jin,  Kyoungsuk and Sim,  Uk},
  year = {2025},
  month = oct,
  pages = {100930}
}

@article{Iyyappan2026,
  title = {Entropically Stabilized Compositionally Complex Prussian Blue Analogues in Electrochemical Energy Storage and Catalytic Applications},
  ISSN = {1614-6840},
  url = {http://dx.doi.org/10.1002/aenm.202505809},
  DOI = {10.1002/aenm.202505809},
  journal = {Advanced Energy Materials},
  publisher = {Wiley},
  author = {Iyyappan,  Madakannu and Senthil,  Chenrayan and Ghosh,  Debasis},
  year = {2026},
  month = jan 
}

@article{Abbaspour2005,
  title = {Electrochemical formation of Prussian blue films with a single ferricyanide solution on gold electrode},
  volume = {584},
  ISSN = {1572-6657},
  url = {http://dx.doi.org/10.1016/j.jelechem.2005.07.008},
  DOI = {10.1016/j.jelechem.2005.07.008},
  number = {2},
  journal = {Journal of Electroanalytical Chemistry},
  publisher = {Elsevier BV},
  author = {Abbaspour,  Abdolkarim and Kamyabi,  Mohammad Ali},
  year = {2005},
  month = oct,
  pages = {117--123}
}

@article{Tharuman2025,
  title = {Electrochemical sensing and catalytic water processing using CoFe-PBA: Experimental and DFT insights},
  volume = {71},
  ISSN = {2214-7144},
  url = {http://dx.doi.org/10.1016/j.jwpe.2025.107301},
  DOI = {10.1016/j.jwpe.2025.107301},
  journal = {Journal of Water Process Engineering},
  publisher = {Elsevier BV},
  author = {Tharuman,  Sharmila and Nataraj,  Nandini and Chen,  Shen-Ming and Vajeeston,  Ponniah and Vellaichamy,  Balakumar},
  year = {2025},
  month = mar,
  pages = {107301}
}

@article{Moretti2018,
  title = {Raman spectroscopy of the photosensitive pigment Prussian blue},
  volume = {49},
  ISSN = {1097-4555},
  url = {http://dx.doi.org/10.1002/jrs.5366},
  DOI = {10.1002/jrs.5366},
  number = {7},
  journal = {Journal of Raman Spectroscopy},
  publisher = {Wiley},
  author = {Moretti,  Giulia and Gervais,  Claire},
  year = {2018},
  month = apr,
  pages = {1198--1204}
}

@article{Weidinger2010,
  title = {Vibrational dynamics of metal cyanides},
  volume = {489},
  ISSN = {0009-2614},
  url = {http://dx.doi.org/10.1016/j.cplett.2010.02.070},
  DOI = {10.1016/j.cplett.2010.02.070},
  number = {4--6},
  journal = {Chemical Physics Letters},
  publisher = {Elsevier BV},
  author = {Weidinger,  Daniel and Sando,  Gerald M. and Owrutsky,  Jeffrey C.},
  year = {2010},
  month = apr,
  pages = {169--174}
}

@article{Ojwang2021,
  title = {Moisture-Driven Degradation Pathways in Prussian White Cathode Material for Sodium-Ion Batteries},
  volume = {13},
  ISSN = {1944-8252},
  url = {http://dx.doi.org/10.1021/acsami.0c22032},
  DOI = {10.1021/acsami.0c22032},
  number = {8},
  journal = {ACS Applied Materials \& Interfaces},
  publisher = {American Chemical Society (ACS)},
  author = {Ojwang,  Dickson O. and Svensson,  Mikael and Njel,  Christian and Mogensen,  Ronnie and Menon,  Ashok S. and Ericsson,  Tore and H\"{a}ggstr\"{o}m,  Lennart and Maibach,  Julia and Brant,  William R.},
  year = {2021},
  month = feb,
  pages = {10054--10063}
}

@article{Sada2024,
  title = {Unveiling the Influence of Water Molecules on the Structural Dynamics of Prussian Blue Analogues},
  volume = {20},
  ISSN = {1613-6829},
  url = {http://dx.doi.org/10.1002/smll.202406853},
  DOI = {10.1002/smll.202406853},
  number = {50},
  journal = {Small},
  publisher = {Wiley},
  author = {Sada,  Krishnakanth and Greene,  Samuel M. and Kmiec,  Steven and Siegel,  Donald J. and Manthiram,  Arumugam},
  year = {2024},
  month = sep 
}

@article{Yun2021,
  title = {Copper Hexacyanoferrate Thin Film Deposition and Its Application to a New Method for Diffusion Coefficient Measurement},
  volume = {11},
  ISSN = {2079-4991},
  url = {http://dx.doi.org/10.3390/nano11071860},
  DOI = {10.3390/nano11071860},
  number = {7},
  journal = {Nanomaterials},
  publisher = {MDPI AG},
  author = {Yun,  Jeonghun and Kim,  Yeongae and Gao,  Caitian and Kim,  Moobum and Lee,  Jae Yoon and Lee,  Chul-Ho and Bae,  Tae-Hyun and Lee,  Seok Woo},
  year = {2021},
  month = jul,
  pages = {1860}
}

@article{Steeger2025,
  title = {Controlling the Morphology and Electrochemical Properties of Electrodeposited Nickel Hexacyanoferrate},
  volume = {12},
  ISSN = {2196-0216},
  url = {http://dx.doi.org/10.1002/celc.202500073},
  DOI = {10.1002/celc.202500073},
  number = {13},
  journal = {ChemElectroChem},
  publisher = {Wiley},
  author = {Steeger,  Tim and Streng,  Raphael L. and Senyshyn,  Anatoliy and Dyadkin,  Vadim and Lamprecht,  Xaver and List,  Roman and Bandarenka,  Aliaksandr S.},
  year = {2025},
  month = apr 
}

@article{Avila2025,
  title = {Variability analysis in memristors based on electrodeposited prussian blue},
  volume = {300},
  ISSN = {0167-9317},
  url = {http://dx.doi.org/10.1016/j.mee.2025.112376},
  DOI = {10.1016/j.mee.2025.112376},
  journal = {Microelectronic Engineering},
  publisher = {Elsevier BV},
  author = {Avila,  L.B. and Cantudo,  A. and Villena,  M.A. and Maldonado,  D. and Araujo,  F. Abreu and M\"{u}ller,  C.K. and Rold\'{a}n,  J.B.},
  year = {2025},
  month = nov,
  pages = {112376}
}

@article{Nordstrand2021,
  title = {Ladder Mechanisms of Ion Transport in Prussian Blue Analogues},
  volume = {14},
  ISSN = {1944-8252},
  url = {http://dx.doi.org/10.1021/acsami.1c20910},
  DOI = {10.1021/acsami.1c20910},
  number = {1},
  journal = {ACS Applied Materials \& Interfaces},
  publisher = {American Chemical Society (ACS)},
  author = {Nordstrand,  Johan and Toledo-Carrillo,  Esteban and Vafakhah,  Sareh and Guo,  Lu and Yang,  Hui Ying and Kloo,  Lars and Dutta,  Joydeep},
  year = {2021},
  month = dec,
  pages = {1102--1113}
}

@article{Pohlitz2022,
  title = {Study of Growth and Properties of Electrodeposited Sodium Iron Hexacyanoferrate Films},
  volume = {15},
  ISSN = {1996-1944},
  url = {http://dx.doi.org/10.3390/ma15217491},
  DOI = {10.3390/ma15217491},
  number = {21},
  journal = {Materials},
  publisher = {MDPI AG},
  author = {Pohlitz,  Michael and M\"{u}ller,  Christian K.},
  year = {2022},
  month = oct,
  pages = {7491}
}

@article{Neff1978,
  author  = {Neff, V. D.},
  title   = {Electrochemical oxidation and reduction of thin films of Prussian Blue},
  journal = {Journal of The Electrochemical Society},
  volume  = {125},
  number  = {6},
  pages   = {886--887},
  year    = {1978},
  doi     = {10.1149/1.2131540}
}

@article{Itaya1982,
  author  = {Itaya, K. and Ataka, T. and Toshima, S.},
  title   = {Spectroelectrochemistry and electrochemical preparation method of Prussian Blue modified electrodes},
  journal = {Journal of the American Chemical Society},
  volume  = {104},
  number  = {18},
  pages   = {4767--4772},
  year    = {1982},
  doi     = {10.1021/ja00383a001}
}

@article{Buser1977,
  author  = {Buser, H. J. and Schwarzenbach, D. and Petter, W. and Ludi, A.},
  title   = {The crystal structure of Prussian Blue: Fe4[Fe(CN)6]3\ensuremath{\cdot}xH2O},
  journal = {Inorganic Chemistry},
  volume  = {16},
  number  = {11},
  pages   = {2704--2710},
  year    = {1977},
  doi     = {10.1021/ic50177a016}
}

@article{Wessells2011,
  author  = {Wessells, C. D. and Peddada, S. V. and Huggins, R. A. and Cui, Y.},
  title   = {Electrochemical intercalation of sodium into Prussian Blue analogues},
  journal = {Nature Communications},
  volume  = {2},
  pages   = {550},
  year    = {2011},
  doi     = {10.1038/ncomms1553}
}

@article{Pillai2024,
  title = {Electrochemical and transport properties of nickel hexacyanoferrate (KNi[Fe(CN)6]) as potential cathode material for aqueous potassium-ion batteries},
  volume = {844},
  ISSN = {0009-2614},
  url = {http://dx.doi.org/10.1016/j.cplett.2024.141273},
  DOI = {10.1016/j.cplett.2024.141273},
  journal = {Chemical Physics Letters},
  publisher = {Elsevier BV},
  author = {Pillai,  Shreeram and Dagadkhair,  Krishna and Salame,  Paresh H.},
  year = {2024},
  month = jun,
  pages = {141273}
}

@article{Xu2018,
  title = {Structure Distortion Induced Monoclinic Nickel Hexacyanoferrate as High-Performance Cathode for Na-Ion Batteries},
  volume = {9},
  ISSN = {1614-6840},
  url = {http://dx.doi.org/10.1002/aenm.201803158},
  DOI = {10.1002/aenm.201803158},
  number = {4},
  journal = {Advanced Energy Materials},
  publisher = {Wiley},
  author = {Xu,  Yue and Wan,  Jing and Huang,  Li and Ou,  Mingyang and Fan,  Chenyang and Wei,  Peng and Peng,  Jian and Liu,  Yi and Qiu,  Yuegang and Sun,  Xueping and Fang,  Chun and Li,  Qing and Han,  Jiantao and Huang,  Yunhui and Alonso,  Jos\'{e} Antonio and Zhao,  Yusheng},
  year = {2018},
  month = dec 
}

@article{Steen2002,
  title = {Structure of Cathodically Deposited Nickel Hexacyanoferrate Thin Films Using XRD and EXAFS},
  volume = {18},
  ISSN = {1520-5827},
  url = {http://dx.doi.org/10.1021/la020352e},
  DOI = {10.1021/la020352e},
  number = {20},
  journal = {Langmuir},
  publisher = {American Chemical Society (ACS)},
  author = {Steen,  William A. and Han,  Sang-Wook and Yu,  Qiuming and Gordon,  Robert A. and Cross,  Julie Olmsted and Stern,  Edward A. and Seidler,  Gerald T. and Jeerage,  Kavita M. and Schwartz,  Daniel T.},
  year = {2002},
  month = aug,
  pages = {7714--7721}
}

@article{Jeerage2002,
  title = {Correlating Nanoscale Structure with Ion Intercalation in Electrodeposited Nickel Hexacyanoferrate Thin Films},
  volume = {14},
  ISSN = {1520-5002},
  url = {http://dx.doi.org/10.1021/cm010156d},
  DOI = {10.1021/cm010156d},
  number = {2},
  journal = {Chemistry of Materials},
  publisher = {American Chemical Society (ACS)},
  author = {Jeerage,  Kavita M. and Steen,  William A. and Schwartz,  Daniel T.},
  year = {2002},
  month = jan,
  pages = {530--535}
}

@article{Liu2024,
  title = {The effect of coprecipitation and heating temperature on structural evolution and electrochemical performance of iron-based prussian blue analogs},
  volume = {28},
  ISSN = {1433-0768},
  url = {http://dx.doi.org/10.1007/s10008-024-06005-2},
  DOI = {10.1007/s10008-024-06005-2},
  number = {11},
  journal = {Journal of Solid State Electrochemistry},
  publisher = {Springer Science and Business Media LLC},
  author = {Liu,  Yuqing and Chu,  Wencheng and Xu,  Yaozu and Yuan,  Zijian and Liu,  Jiarui and Zhao,  Haitao and Liu,  Quan and Zhang,  Wu},
  year = {2024},
  month = jul,
  pages = {4105--4118}
}

@article{Yang2016,
  title = {Influence of Structural Imperfection on Electrochemical Behavior of Prussian Blue Cathode Materials for Sodium Ion Batteries},
  volume = {163},
  ISSN = {1945-7111},
  url = {http://dx.doi.org/10.1149/2.0031610jes},
  DOI = {10.1149/2.0031610jes},
  number = {9},
  journal = {Journal of The Electrochemical Society},
  publisher = {The Electrochemical Society},
  author = {Yang,  Yang and Liu,  Ershuai and Yan,  Xiaomin and Ma,  Chunrong and Wen,  Wen and Liao,  Xiao-Zhen and Ma,  Zi-Feng},
  year = {2016},
  pages = {A2117--A2123}
}

@article{Kettle2011,
  author  = {Kettle, S. F. A. and Diana, E. and Marchese, E. M. C. and Boccaleri, E.},
  title   = {The vibrational spectra of the cyanide ligand revisited: The $\nu$(CN) infrared and Raman spectroscopy of Prussian blue and its analogues},
  journal = {Journal of Raman Spectroscopy},
  volume  = {42},
  number  = {11},
  pages   = {2006--2014},
  year    = {2011},
  doi     = {10.1002/jrs.2944}
}

@article{Javadi2026,
  title = {Prussian blue analogues for the next-generation of beyond lithium energy storage},
  volume = {141},
  ISSN = {2352-152X},
  url = {http://dx.doi.org/10.1016/j.est.2025.119244},
  DOI = {10.1016/j.est.2025.119244},
  journal = {Journal of Energy Storage},
  publisher = {Elsevier BV},
  author = {Javadi,  Omid and Quick,  Shiva and Rastegar,  Seyed Omid},
  year = {2026},
  month = jan,
  pages = {119244}
}

@article{Duan2026,
  title = {Simultaneous detection of tetracycline and kanamycin by electrochemical sensors based on dual-signal probes Au/Mn-based Prussian blue analogs and methylene blue-functionalized AuPt NFs},
  volume = {448},
  ISSN = {0925-4005},
  url = {http://dx.doi.org/10.1016/j.snb.2025.138933},
  DOI = {10.1016/j.snb.2025.138933},
  journal = {Sensors and Actuators B: Chemical},
  publisher = {Elsevier BV},
  author = {Duan,  Linjing and Suo,  Zhiguang and Yan,  Yaqi and Zhang,  Jinmin and Guo,  Rui and Wei,  Min and Jin,  Huali and Zhao,  Renyong},
  year = {2026},
  month = feb,
  pages = {138933}
}

@article{Singh2026,
  title = {Recent advances and perspectives on Prussian blue analogue-based catalysts for electrocatalytic small-molecule oxidation reactions},
  volume = {552},
  ISSN = {0010-8545},
  url = {http://dx.doi.org/10.1016/j.ccr.2025.217546},
  DOI = {10.1016/j.ccr.2025.217546},
  journal = {Coordination Chemistry Reviews},
  publisher = {Elsevier BV},
  author = {Singh,  Baghendra},
  year = {2026},
  month = apr,
  pages = {217546}
}

@article{Wu2024,
  title = {Vacancy Remediation in Prussian Blue Analogs for High-Performance Sodium and Potassium Ion Batteries},
  volume = {35},
  ISSN = {1616-3028},
  url = {http://dx.doi.org/10.1002/adfm.202418018},
  DOI = {10.1002/adfm.202418018},
  number = {13},
  journal = {Advanced Functional Materials},
  publisher = {Wiley},
  author = {Wu,  Ruixue and Ren,  Bo and Wang,  Xianda and Lin,  Jie and Li,  Xiaoxia and Zheng,  Jinlong and Yang,  Hui Ying and Shang,  Yang},
  year = {2024},
  month = dec 
}

@book{Cullity2001,
  author    = {Cullity, Bernard D. and Stock, Stuart R.},
  title     = {Elements of X-Ray Diffraction},
  publisher = {Prentice Hall},
  edition   = {3},
  year      = {2001},
  address   = {Upper Saddle River, NJ}
}

@article{Lee2024,
  title = {Effect of the Interaction between Transition Metal Redox Center and Cyanide Ligand on Structural Evolution in Prussian White Cathodes},
  volume = {18},
  ISSN = {1936-086X},
  url = {http://dx.doi.org/10.1021/acsnano.3c08271},
  DOI = {10.1021/acsnano.3c08271},
  number = {3},
  journal = {ACS Nano},
  publisher = {American Chemical Society (ACS)},
  author = {Lee,  Ju-Hyeon and Bae,  Jin-Gyu and Kim,  Min Sung and Heo,  Jeong Yeon and Lee,  Hyeon Jeong and Lee,  Ji Hoon},
  year = {2024},
  month = jan,
  pages = {1995--2005}
}

@book{Waseda1980,
  author    = {Waseda, Yoshio},
  title     = {The Structure of Non-Crystalline Materials: Liquids and Amorphous Solids},
  publisher = {McGraw-Hill},
  address   = {New York},
  year      = {1980},
  isbn      = {9780070699700}
}

@book{Warren1969,
  author = {Warren, B. E.},
  title = {X-Ray Diffraction},
  publisher = {Addison-Wesley},
  year = {1969}
}

@article{Widmann2002,
  author = {Widmann, A. and Kahlert, H. and Petrovic-Prelevic, I. and Wulff, H. and Yakhmi, J. V. and Scholz, F.},
  title = {Structure, Thermodynamics, and Kinetics of Prussian Blue},
  journal = {Langmuir},
  volume = {18},
  pages = {3830--3840},
  year = {2002}
}

@book{Young1993,
  author = {Young, R. A.},
  title = {The Rietveld Method},
  publisher = {Oxford University Press},
  year = {1993}
}

@book{Nakamoto2009,
  author = {Nakamoto, K.},
  title = {Infrared and Raman Spectra of Inorganic and Coordination Compounds, Part B: Applications in Coordination, Organometallic, and Bioinorganic Chemistry},
  publisher = {Wiley},
  year = {2009},
  edition = {6th}
}

@article{Nielsen2026,
  author  = {Nielsen, Ida and Eremenko, Maksim and Zhang, Yuanpeng and Tucker, Matthew G. and Brant, William R.},
  title   = {Local structure of hydrated and dehydrated Prussian white cathode materials},
  journal = {Journal of Materials Chemistry C},
  year    = {2026},
  doi     = {10.1039/D5TC03143E},
  note    = {Advance article}
}

@article{Jain2024,
  title = {Deciphering the Synergistic Effect in Strong Metal--Support Interaction toward Surface Reconstruction of Prussian Blue Analogue-Based Pre-Catalysts for Boosted Bifunctional Oxygen Electrocatalysis via Operando Raman Spectroscopy},
  volume = {128},
  ISSN = {1932-7455},
  url = {http://dx.doi.org/10.1021/acs.jpcc.4c04479},
  DOI = {10.1021/acs.jpcc.4c04479},
  number = {39},
  journal = {The Journal of Physical Chemistry C},
  publisher = {American Chemical Society (ACS)},
  author = {Jain,  Priya and Ingole,  Pravin Popinand},
  year = {2024},
  month = sep,
  pages = {16380--16392}
}

@article{Avila2020,
  author  = {Avila, L. B. and M{\"u}ller, C. K. and Hildebrand, D. and Faita, F. L. and Baggio, B. F. and Cid, C. C. P. and Pasa, A. A.},
  title   = {Resistive Switching in Electrodeposited Prussian Blue Layers},
  journal = {Materials},
  year    = {2020},
  volume  = {13},
  pages   = {5618},
  doi     = {10.3390/ma13245618}
}

@article{Faita2022,
  author  = {Faita, F. L. and Avila, L. B. and Silva, J. P. B. and Boratto, M. H. and Pl{\'a} Cid, C. C. and Graeff, C. F. O. and Gomes, M. J. M. and M{\"u}ller, C. K. and Pasa, A. A.},
  title   = {Abnormal Resistive Switching in Electrodeposited Prussian White Thin Films},
  journal = {Journal of Alloys and Compounds},
  year    = {2022},
  volume  = {896},
  pages   = {162971},
  doi     = {10.1016/j.jallcom.2021.162971}
}

@article{Avila2022,
  author  = {Avila, L. B. and Serrano Arambulo, P. C. and Dantas, A. and Cuevas-Arizaca, E. E. and Kumar, D. and M{\"u}ller, C. K.},
  title   = {Study on the Electrical Conduction Mechanism of Unipolar Resistive Switching Prussian White Thin Films},
  journal = {Nanomaterials},
  year    = {2022},
  volume  = {12},
  pages   = {2881},
  doi     = {10.3390/nano12162881}
}

@article{Avila2024PB,
  author  = {Avila, L. B. and Serrano, P. A. and Quispe, L. T. and Dantas, A. and Costa, D. P. and Arizaca, E. E. C. and Ch{\'a}vez, D. P. P. and Portugal, C. D. V. and M{\"u}ller, C. K.},
  title   = {Prussian Blue Anchored on Reduced Graphene Oxide Substrate Achieving High Voltage in Symmetric Supercapacitor},
  journal = {Materials},
  year    = {2024},
  volume  = {17},
  pages   = {3782},
  doi     = {10.3390/ma17153782}
}

@article{Avila2024Perylene,
  author  = {Avila, L. B. and Chulkin, P. and Serrano, P. A. and Dreyer, J. P. and Berteau-Rainville, M. and Orgiu, E. and Fran{\c{c}}a, L. D. L. and Zimmermann, L. M. and Bock, H. and Faria, G. C. and Eccher, J. and Bechtold, I. H.},
  title   = {Perylene-Based Columnar Liquid Crystal: Revealing Resistive Switching for Nonvolatile Memory Devices},
  journal = {Journal of Molecular Liquids},
  year    = {2024},
  volume  = {402},
  pages   = {124757},
  doi     = {10.1016/j.molliq.2024.124757}
}

@article{Quispe2021,
  author  = {Quispe, L. T. and Avila, L. B. and Linhares, A. A. and others},
  title   = {p-Type NiO Thin Films Obtained via an Electrochemical-Thermal Route},
  journal = {Journal of Materials Science: Materials in Electronics},
  year    = {2021},
  volume  = {32},
  pages   = {5372--5380},
  doi     = {10.1007/s10854-021-05260-7}
}

\end{document}